\documentclass[conference,a4paper]{IEEEtran}
\IEEEoverridecommandlockouts

\usepackage{cite}
\usepackage{amsmath,amssymb,amsfonts}
\usepackage{algorithmic}
\usepackage{graphicx}
\usepackage{textcomp}
\usepackage{xcolor}
\usepackage{booktabs}

\def\BibTeX{{\rm B\kern-.05em{\sc i\kern-.025em b}\kern-.08em
    T\kern-.1667em\lower.7ex\hbox{E}\kern-.125emX}}
\begin{document}

\title{Adaptive Sampling for Spatiotemporal Anomaly Monitoring in Wireless Sensor Networks}

\author{%
\IEEEauthorblockN{Guoqing~Lu, Yixuan~Sun, Yiwen~Jiang, and
Bernard~Butler\textsuperscript{*}}
\IEEEauthorblockA{South East Technological University, Waterford, Ireland}
\thanks{\textsuperscript{*}Corresponding author: bbutler@ieee.org}
}
\maketitle

\begin{abstract}
 Long-term environmental monitoring in wireless sensor networks (WSNs) often uses sparse sampling to extend network lifetime, but sparse sensing can miss short-lived, localized, and potentially diffusive anomalies. This paper proposes a sentinel-assisted adaptive sampling framework as a cooperative sensing-control pipeline for WSN anomaly monitoring. During normal periods, nodes perform sparse sensing driven by Kalman filter (KF) predictive uncertainty. During anomalous periods, continuously sampled sentinel nodes perform hybrid GLR-based detection with node-relative thresholds, and local detections trigger one-hop neighborhood wake-up with recovery-aware alert control.

Experiments on the Intel Berkeley Research Lab temperature dataset with abrupt random spatiotemporal anomalies show that the proposed method raises the anomaly-window sampling ratio (AWSR) from 0.439 to 0.933 in the main experiment. It also improves AWSR over Adaptive Data Acquisition with Energy Efficiency and Critical-Sensing Guarantee (AAS) and Adapted e-Sampling while reducing total cost by 15.4\% and 2.1\%, respectively. These results show that integrating KF-based sparse sampling, sentinel GLR detection, and local alert propagation improves anomaly-window visibility while maintaining a lower sampling-cost trade-off.
\end{abstract}

\section{Introduction}
\label{sec:intro}
Wireless sensor networks (WSNs) are expected to operate for long periods in applications such as environmental monitoring, smart agriculture, structural health monitoring, and industrial Internet of Things, where node-level energy budgets constrain system availability \cite{akyildiz2002wsn,yu2024wssn}. Persistent high-frequency sensing is wasteful during stationary periods, while overly sparse sampling can miss abrupt events. Balancing energy efficiency and event awareness is therefore a key sensing-layer issue.

Existing work has studied fixed-rate sampling, event-triggered sensing, adaptive sampling, and state-estimation-based control \cite{tian2025wsnstateestimation,ahmad2024oakf}. Prediction-uncertainty-driven methods are effective under stationary conditions, but their decisions are often dominated by node-local temporal information. For spatially localized or diffusive anomalies, purely local control may still cause delayed detection, local misses, and weak regional response.

This paper proposes a sentinel-assisted adaptive sampling framework. During normal periods, nodes use low-cost sampling driven by Kalman filter (KF) predictive uncertainty \cite{kalman1960}. During anomalous periods, a small subset of continuously sampled \emph{sentinel nodes} applies a two-sided Generalized Likelihood Ratio (GLR) detector with node-relative elevation thresholds. Local detections then trigger one-hop neighbor wake-up, while recovery-aware alert control avoids persistent high-rate sensing. The contribution lies in integrating these components into a cooperative WSN sensing-control pipeline that shifts sensing effort from normal-period sparsity to anomaly-window regional observation.

Beyond online detection and sampling control, this paper explicitly distinguishes between the spatial graph and the communication graph. The former describes anomaly influence structure and sentinel spatial coverage, whereas the latter defines the reachability of control propagation. Based on this framework, the proposed method further adopts spatial coverage optimization to select sentinel deployment locations so as to improve regional visibility under a limited budget of continuously sampled nodes \cite{krause2008sensorplacement}.

The evaluation in this paper considers event-level detection, regional coverage, anomaly-window sensing intensity, and overall cost. Section~\ref{subsec:exp1} compares the proposed mechanism with Full sampling, Pure KF sparse, and Oracle-local wake-up, while Section~\ref{subsec:exp2} compares it with AAS and Adapted e-Sampling under a unified comparison runner \cite{rao2019aas,bhuiyan2017esampling}.
The main contributions are:
\begin{itemize}
    \item A cooperative sensing-control pipeline integrating KF uncertainty-driven sparse sampling, sentinel hybrid GLR detection, one-hop alert propagation, and recovery-aware local control for abrupt anomaly monitoring in WSNs.
    \item A spatially optimized sentinel deployment strategy for improving regional visibility under a limited continuously sampled-node budget.
    \item A local-evidence-driven neighborhood wake-up mechanism that improves anomaly-window observation while reducing energy wastage by avoiding unnecessary persistent and widespread high-rate sensing and communication outside affected regions and during normal periods.
\end{itemize}

\section{Related Work}
\label{sec:related}
Existing sampling optimization in WSNs mainly targets either long-term reduction of redundant sensing and transmission or timely awareness of critical events under energy constraints. The problem addressed in this paper is closer to the latter, while still borrowing correlation-driven energy-saving ideas during normal periods.

Correlation-driven adaptive sampling methods reduce sensing and transmission cost by exploiting temporal or spatiotemporal correlations among sensor readings \cite{karaki2019adaptive,cai2018stcsta}. However, their control is typically organized at the round or cluster level and is therefore more suitable for relatively stationary monitoring tasks than for short-lived, spatially localized abrupt anomalies.

AAS emphasizes critical-data assurance by switching between energy-saving and prediction-driven acquisition modes according to local data conditions \cite{rao2019aas}. Although it also balances energy efficiency and event awareness, it mainly relies on per-node mode switching around local thresholds and does not explicitly provide neighborhood-level collaborative response to locally diffusive anomalies.

e-Sampling is an event-sensitive autonomous sensing scheme that adjusts sensing frequency according to local signal dynamics and event indications \cite{bhuiyan2017esampling}. Its event discrimination mainly depends on frequency changes in individual node signals, making it more suitable for high-frequency dynamic events than for locally abrupt and spatially diffusive anomalies over slowly varying backgrounds. In contrast, the present setting requires local detections to be quickly translated into neighborhood-level supplemental sensing.

Overall, existing work has shown the value of dynamic sampling control for reducing WSN cost, but most methods do not explicitly address how to increase observation density around a localized anomaly at low cost after the anomaly emerges. The proposed method addresses this gap by combining KF-based sparse sensing with sentinel-triggered local wake-up, AAS and Adapted e-Sampling are therefore used as the main comparison baselines.

\section{Proposed Method}
\label{sec:method}
This section presents the proposed sentinel-assisted adaptive sampling method through five components: state prediction, anomaly modeling, sentinel detection, alert propagation, and sentinel placement optimization.

\subsection{Network and State Model}
\label{subsec:network}
Consider a wireless sensor network (WSN) consisting of $N$ nodes, with node set $\mathcal{V}=\{1,\ldots,N\}$ and discrete time steps $k=1,\ldots,T$. The WSN collects temperature measurements. Nodes are divided into ordinary nodes and sentinel nodes: ordinary nodes mainly follow uncertainty-driven adaptive sampling, whereas sentinel nodes keep sampling continuously and serve as the primary trigger source for anomaly detection and local wake-up.

The state vector of node $i$ at time step $k$ is defined as
\begin{equation}
x_k^{(i)}=
\begin{bmatrix}
\theta_k^{(i)}\\
\dot{\theta}_k^{(i)}
\end{bmatrix},
\end{equation}
where $\theta_k^{(i)}$ denotes the physical ambient temperature measured at node $i$, rather than an alert-level or node-activity indicator, and $\dot{\theta}_k^{(i)}$ denotes its temperature variation rate. The observation is denoted by $z_k^{(i)}$. 

\subsection{KF-Based Prediction and Adaptive Sampling}
\label{subsec:kf}
All nodes share the same two-dimensional linear Kalman filter (KF) backbone for state estimation and uncertainty tracking \cite{kalman1960}.

The state-transition and observation models of each node are written as
\begin{align}
x_k^{(i)} &= F x_{k-1}^{(i)} + w_k^{(i)},\\
z_k^{(i)} &= H x_k^{(i)} + v_k^{(i)},
\end{align}
where $F$ and $H$ denote the state-transition and observation matrices, and $w_k^{(i)}\sim\mathcal{N}(0,Q)$ and $v_k^{(i)}\sim\mathcal{N}(0,R)$ denote the process and measurement noise, respectively. The standard prediction step is
\begin{align}
\hat{x}_{k|k-1}^{(i)} &= F\hat{x}_{k-1|k-1}^{(i)},\\
P_{k|k-1}^{(i)} &= FP_{k-1|k-1}^{(i)}F^T+Q,
\end{align}
When a measurement is available, the update uses
\begin{align}
\nu_k^{(i)} &= z_k^{(i)}-H\hat{x}_{k|k-1}^{(i)},\\
S_k^{(i)} &= HP_{k|k-1}^{(i)}H^T+R,
\end{align}
followed by the Kalman gain
\begin{equation}
K_k^{(i)} = P_{k|k-1}^{(i)}H^TS_k^{(i)-1},
\end{equation}
and the posterior update
\begin{align}
\hat{x}_{k|k}^{(i)} &= \hat{x}_{k|k-1}^{(i)} + K_k^{(i)}\nu_k^{(i)},\\
P_{k|k}^{(i)} &= (I-K_k^{(i)}H)P_{k|k-1}^{(i)}(I-K_k^{(i)}H)^T + K_k^{(i)}RK_k^{(i)T},
\end{align}
where $I$ is the identity matrix, and the Joseph form is adopted for covariance updating to improve numerical stability~\cite{joseph1968}.

Ordinary nodes do not sample at every step in the energy-saving mode. Instead, a measurement is triggered according to the predictive observation uncertainty
\begin{equation}
\sigma_{y,k}^{2(i)} = HP_{k|k-1}^{(i)}H^T.
\end{equation}
When
\begin{equation}
\sigma_{y,k}^{2(i)} > \gamma,
\end{equation}
node $i$ performs sampling; otherwise, it keeps the predicted state. Here, $\gamma$ is the uncertainty threshold. Sentinel nodes do not follow this sparse rule and instead sample continuously so that the downstream detector always receives uninterrupted inputs.

\subsection{Anomaly Modeling}
\label{subsec:anomaly}
To evaluate anomaly detection and local response under shared background data, controllable anomalies are superimposed on real temperature sequences. Let $\tilde{\theta}_k^{(i)}$ denote the anomaly-corrupted true temperature of node $i$ at time $k$. Then
\begin{equation}
\tilde{\theta}_k^{(i)}=\theta_k^{(i)}+\sum_{e\in\mathcal{E}}\Delta_{e,k}^{(i)},
\end{equation}
where $\mathcal{E}$ is the set of injected events and $\Delta_{e,k}^{(i)}$ is the additive deviation introduced by event $e$ to node $i$ at time step $k$.

The first class of anomalies is static spatial anomalies. For event $e$, let $c_e$ be the center node, $t_e^0$ the starting time, $D_e$ the duration in time steps, $r_e$ the influence radius, and $A_e$ the base amplitude. If $d(i,c_e)$ denotes the spatial distance between node $i$ and the center node, the static spatial anomaly is written as
\begin{equation}
\Delta_{e,k}^{(i)}=
\begin{cases}
A_e\left(1-\dfrac{d(i,c_e)}{r_e}\right), &
\begin{aligned}[t]
&t_e^0 \le k < t_e^0+D_e,\\
&d(i,c_e)\le r_e,
\end{aligned}\\
0, & \text{otherwise},
\end{cases}
\end{equation}
where the anomaly magnitude decays linearly with distance, representing a stable local disturbance that affects adjacent nodes within a region.

The second class is spatiotemporal diffusion anomalies. Let $s_e$ denote the source node of event $e$, $B_e$ the base amplitude, $H_e$ the maximum diffusion hop count, $\tau_e$ the propagation delay per hop, and $\alpha_e\in(0,1]$ the per-hop attenuation factor. Let $h(i,s_e)$ denote the hop distance from node $i$ to the source node in the propagation structure. The diffusion anomaly is written as
\begin{equation}
\Delta_{e,k}^{(i)}=
\begin{cases}
B_e\alpha_e^{h(i,s_e)}, &
\Psi_{e,k}^{(i)}=1,\\
0, & \text{otherwise},
\end{cases}
\end{equation}
where the indicator $\Psi_{e,k}^{(i)}=1$ means that
\begin{equation}
h(i,s_e)\le H_e,\qquad
t_e^0+h(i,s_e)\tau_e \le k < t_e^0+h(i,s_e)\tau_e + D_e.
\end{equation}

Along the temporal dimension, the main experiment uses abrupt onset and abrupt recovery for both anomaly classes: the offset appears immediately when a node enters the event window and vanishes immediately when the window ends. This controlled setting separates anomaly occurrence, persistence, and recovery so that different detection and local wake-up mechanisms can be compared more directly.

\subsection{Sentinel Detection and Local Wake-Up}
\label{subsec:alert}
\begin{figure}[t]
\centering
\includegraphics[width=0.95\linewidth]{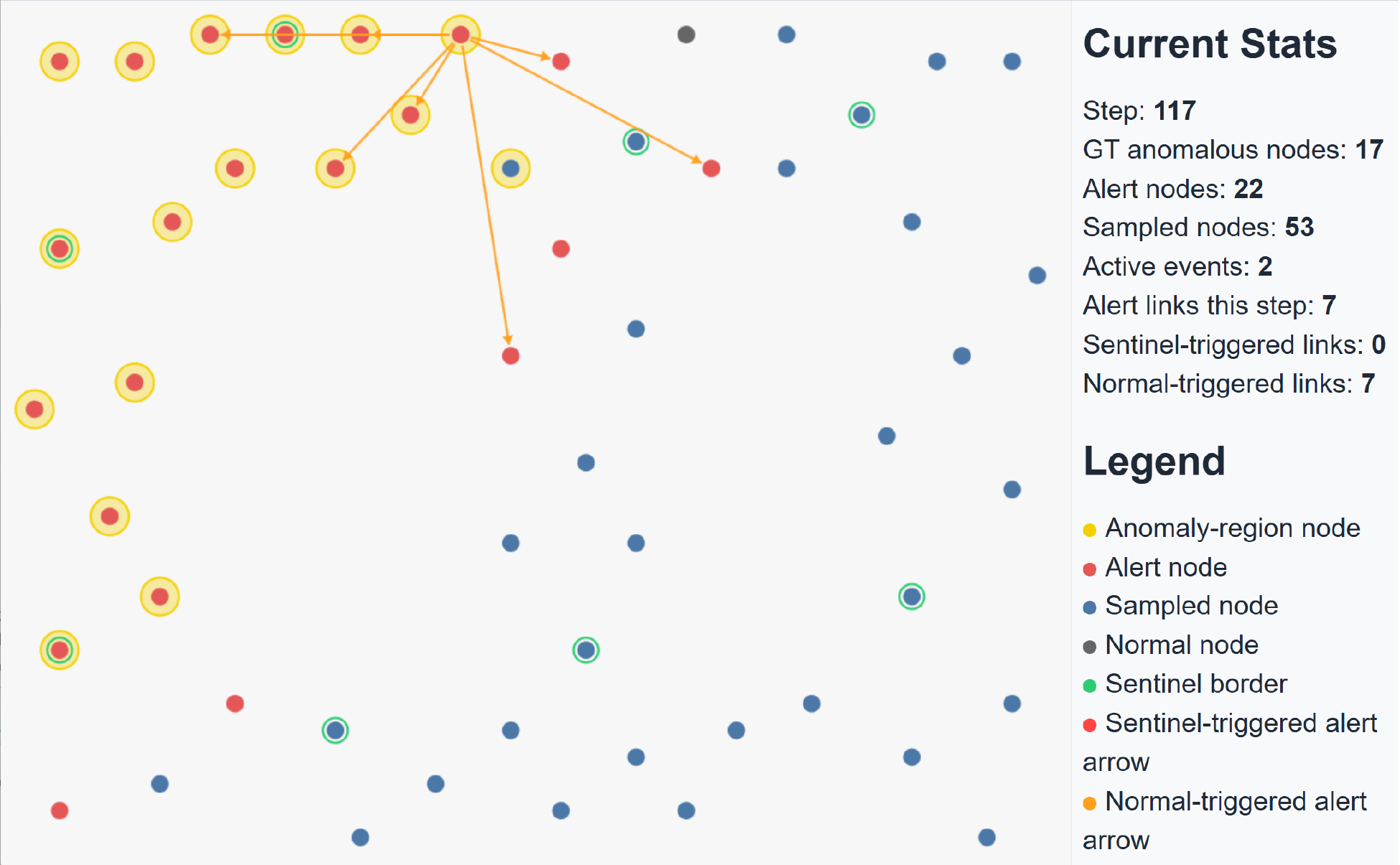}
\caption{Illustration of the node-level alert logic. The figure shows anomaly-region nodes, alert states, actual sampling, and alert propagation, highlighting how local detection is converted into neighborhood wake-up and dense sensing.}
\label{fig:alertlogic}
\end{figure}

All nodes run the KF, but sentinel nodes sample continuously and serve as the main trigger source in the response chain, as illustrated in Fig. \ref{fig:alertlogic}. When a trigger source is judged anomalous, the system propagates a one-hop wake-up signal to nearby nodes; sampled ordinary nodes can also call the same detector when self-triggering is enabled.

The current main experiment adopts a GLR detector built on the signed normalized innovation. For node $i$, the innovation and innovation covariance at time step $k$ are
\begin{align}
\nu_k^{(i)} &= z_k^{(i)}-H\hat{x}_{k|k-1}^{(i)},\\
S_k^{(i)} &= HP_{k|k-1}^{(i)}H^T+R.
\end{align}
Based on them, the direction-aware normalized innovation signal is defined as
\begin{equation}
r_k^{(i)}=\frac{\nu_k^{(i)}}{\sqrt{S_k^{(i)}}}.
\end{equation}
Unlike the normalized innovation squared (NIS), $r_k^{(i)}$ preserves the sign of the deviation and can therefore distinguish between abrupt upward and abrupt downward changes.

For a recent window of length $w$, the positive and negative GLR statistics are defined as
\begin{align}
G_{k,+}^{(i)}(w) &= \frac{\left[\max\left(0,\sum_{t=k-w+1}^{k} r_t^{(i)}\right)\right]^2}{2w},\\
G_{k,-}^{(i)}(w) &= \frac{\left[\max\left(0,-\sum_{t=k-w+1}^{k} r_t^{(i)}\right)\right]^2}{2w}.
\end{align}
Taking the optimum over the window set $\mathcal{W}=\{w_{\min},\ldots,w_{\max}\}$ gives
\begin{align}
\bar{G}_{k,+}^{(i)} &= \max_{w\in\mathcal{W}} G_{k,+}^{(i)}(w),\\
\bar{G}_{k,-}^{(i)} &= \max_{w\in\mathcal{W}} G_{k,-}^{(i)}(w).
\end{align}
The current main experiment uses two-sided triggering:
\begin{equation}
\bar{G}_{k}^{(i)}=\max\left(\bar{G}_{k,+}^{(i)},\bar{G}_{k,-}^{(i)}\right).
\end{equation}
However, the current implementation no longer relies solely on a single global fixed threshold. Instead, it uses a hybrid criterion composed of an absolute threshold and a node-relative elevation threshold. Specifically, for node $i$, the system maintains a recent GLR background sequence under normal, non-alert tracking,
$\mathcal{B}_k^{(i)}=\{\bar{G}_{t}^{(i)}\}$, and constructs a node-adaptive reference based on its median and median absolute deviation (MAD) \cite{rousseeuw1993mad}:
\begin{align}
\mu_k^{(i)} &= \mathrm{median}\left(\mathcal{B}_k^{(i)}\right),\\
\sigma_k^{(i)} &= \max\!\left(1.4826\cdot \mathrm{median}\left(|\mathcal{B}_k^{(i)}-\mu_k^{(i)}|\right),\ \sigma_{\min}\right),
\end{align}
where $\sigma_{\min}$ is a lower bound on the scale to avoid degenerately small thresholds under extremely low background variation. The relative triggering threshold is then defined as
\begin{equation}
\lambda_{\mathrm{rel},k}^{(i)}=\max\left(\lambda_{\mathrm{floor}},\ \mu_k^{(i)}+\kappa \sigma_k^{(i)}\right),
\end{equation}
and the final trigger criterion is
\begin{equation}
\bar{G}_{k}^{(i)} \ge \lambda_{\mathrm{GLR}}
\quad \text{or} \quad
\bar{G}_{k}^{(i)} \ge \lambda_{\mathrm{rel},k}^{(i)}.
\end{equation}
This hybrid trigger combines a global hard threshold with a node-relative elevation threshold so that clearly significant changes remain detectable while node-specific abnormal elevations can also be captured in time. Because the system keeps both $\bar{G}_{k,+}^{(i)}$ and $\bar{G}_{k,-}^{(i)}$, it can further identify whether the dominant current change is upward or downward.

At the triggering stage, a latching mechanism is introduced to avoid repeatedly broadcasting redundant wake-up commands during anomaly persistence: once a node triggers and sends one round of alert messages, it enters a latched state and does not rebroadcast at every subsequent step.

After receiving an alert, an ordinary node passes through a short verification window, confirmed alert tracking, and recovery-aware exit: it returns to sparse sampling if local confirmation is absent, stays at a high sampling rate after confirmed alert, and exits only when the recovery boundary is detected. During confirmed alert tracking, the opposite-direction recovery cliff is used only for exit judgment rather than being treated as a fresh anomaly source.

Alert propagation currently follows a one-hop diffusion policy on the communication graph. Let $\mathrm{dist}_c(i,j)$ denote the communication-hop distance from trigger source node $i$ to node $j$. Then node $j$ is included in the current wake-up set if
\begin{equation}
\mathrm{dist}_c(i,j)\le 1.
\end{equation}
This one-hop propagation mechanism rapidly converts local detections into temporary high-rate sampling over the surrounding neighborhood, thereby improving local visibility within anomaly windows.

\subsection{Sentinel Placement via Spatial Optimal Cover}
\label{subsec:placement}
Since sentinel nodes need to sample continuously, deploying too many of them leads to substantial normal-period energy consumption, while deploying too few creates spatial blind spots. Sentinel placement is modeled as a coverage optimization problem: under a given sentinel budget, the objective is to place sentinels so that as many nodes as possible fall within the effective coverage radius of at least one sentinel. Let the binary decision variable $x_i\in\{0,1\}$ indicate whether node $i$ is selected as a sentinel, and let $y_j\in\{0,1\}$ indicate whether node $j$ is covered. Let $K$ be the maximum number of allowed sentinels, and let $a_{ij}\in\{0,1\}$ indicate whether node $j$ lies within node $i$'s coverage radius $r_{\mathrm{cov}}$. The problem is then formulated as
\begin{align}
\max_{\{x_i\},\{y_j\}} \quad & \sum_{j\in\mathcal{V}} y_j,\\
\text{s.t.}\quad
& y_j \le \sum_{i\in\mathcal{V}} a_{ij}x_i,\quad \forall j\in\mathcal{V},\\
& \sum_{i\in\mathcal{V}} x_i \le K,\\
& x_i\in\{0,1\},\ y_j\in\{0,1\}.
\end{align}
Here, $a_{ij}=1$ means $d(i,j)\le r_{\mathrm{cov}}$, where $d(i,j)$ is the Euclidean distance between nodes.

\section{Experiments}
\label{sec:experiments}
This section contains two numerical experiments with complementary purposes. Experiment 1 compares the proposed method with Full sampling, Pure KF sparse, and Oracle-local wake-up to examine its behavior against reference cases that represent uniform high-rate sensing, low-cost sparse sensing, and idealized local response. Experiment 2 compares the proposed method with AAS and Adapted e-Sampling to evaluate its performance against retained adaptive sampling baselines from the literature. Together, the two experiments separate mechanism-level validation from external baseline comparison, while all methods share the same data window, time axis, anomaly plan, base KF model, and random seed.

\subsection{Simulation Setup}
\label{subsec:setup}
The experiments use the Intel Berkeley Research Lab temperature dataset, retaining 54 sensor nodes over the common window from 2004-02-28 01:03:00 to 05:03:00 with a 31 s step. Because the original observations are not strictly synchronized, the data are aligned and resampled on a shared time axis before simulation.

All nodes share the same two-dimensional KF backbone. Ordinary nodes use variance-threshold sparse sampling, sentinel nodes remain continuously sampled, and the sentinel ratio is fixed at 0.15. The detector uses a two-sided GLR with relative triggering, alerts follow one-hop diffusion with a 5-step verification window, and 10 random anomalies are injected. This controlled injection setting is used to compare methods under the same anomaly plan, data window, KF model, and random seed. Detailed settings are summarized in Table \ref{tab:setup}.

\begin{table}[t]
\centering
\caption{Main experimental settings.}
\label{tab:setup}
\footnotesize
\setlength{\tabcolsep}{6pt}
\renewcommand{\arraystretch}{1.12}
\begin{tabular}{ll}
\toprule
\textbf{Parameter} & \textbf{Setting} \\
\midrule
Time step & 31 s \\
Number of nodes & 54 \\
Sentinel ratio & 0.15 \\
GLR window / absolute threshold & 1--8 / 6.0 \\
Relative GLR trigger & $\kappa = 3.0,\ \lambda_{\mathrm{floor}} = 3.4$ \\
Alert verification wait & 5 steps \\
Number of random events & 10 \\
Magnitude range & 5.0--8.0 \\
Static anomaly radius & 4.0--10.0 \\
Diffusion hop range & 2--5 \\
\bottomrule
\end{tabular}
\end{table}

\subsection{Experiment 1: Main Method Effectiveness}
\label{subsec:exp1}
Experiment 1 evaluates mechanism-level effectiveness by comparing the proposed method with Full sampling, Pure KF sparse, and Oracle-local wake-up.

Six metrics are used: Event Detection Rate (EDR) and Onset Detection Delay (ODD) for event-level awareness, alert coverage and AWSR for regional response during anomaly windows, total energy proxy cost for overall cost, and anomaly-window RMSE for reconstruction quality.

Table \ref{tab:task10main} reports the representative Task10 main results. Compared with Pure KF sparse, Sentinel-assisted KF keeps EDR at 0.900, reduces ODD from 1.778 to 1.444, and improves Coverage and AWSR from 0.130/0.439 to 0.977/0.933, showing the benefit of sentinel-triggered local densification.

Compared with Full sampling, Sentinel-assisted KF still operates at much lower Global Sampling Rate (GSR) and energy cost (0.758 and 298127 versus 1.000 and 392829) while maintaining AWSR of 0.933 and lower anomaly-window RMSE than Pure KF sparse.

Compared with Oracle-local wake-up, Coverage and AWSR remain slightly lower, indicating that the main remaining gap lies in propagation range and regional hit quality.

\begin{figure}[!t]
\centering
\includegraphics[width=\linewidth]{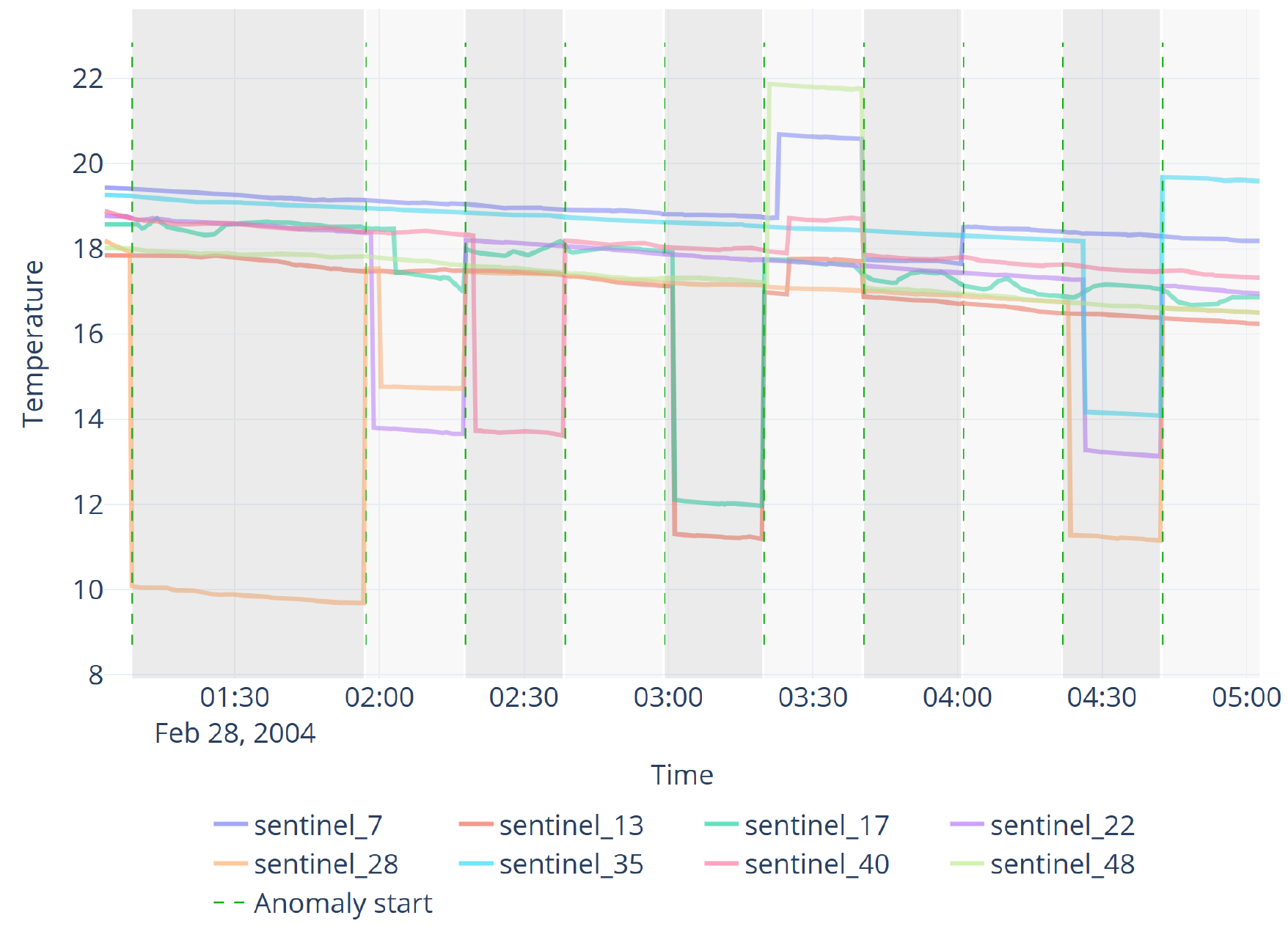}
\caption{Global temperature trajectories of sentinel nodes in the representative main experiment, together with anomaly windows and onset markers.}
\label{fig:globaltemp}
\end{figure}

\begin{figure}[!t]
\centering
\includegraphics[width=0.95\linewidth]{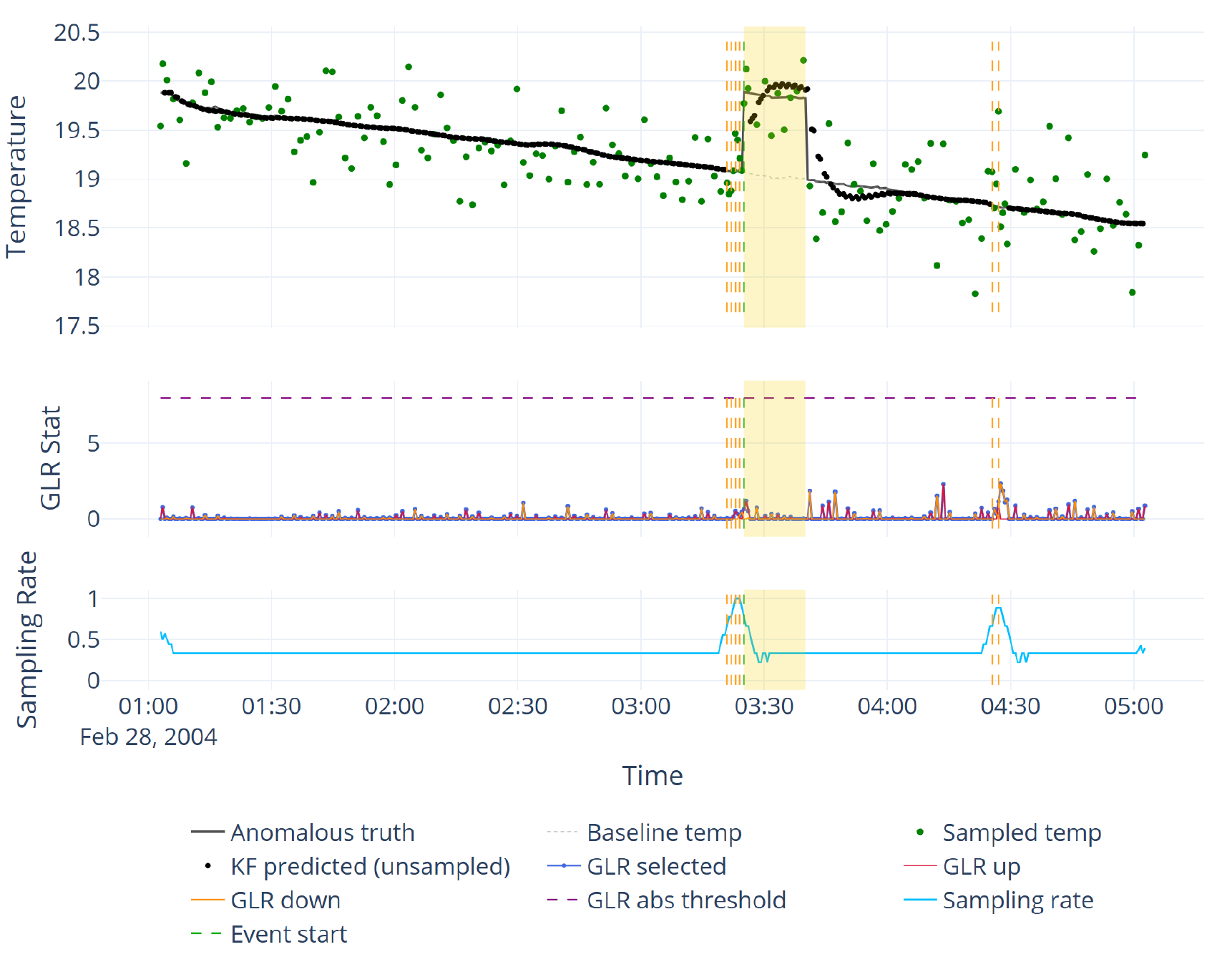}\\[2pt]
\includegraphics[width=0.95\linewidth]{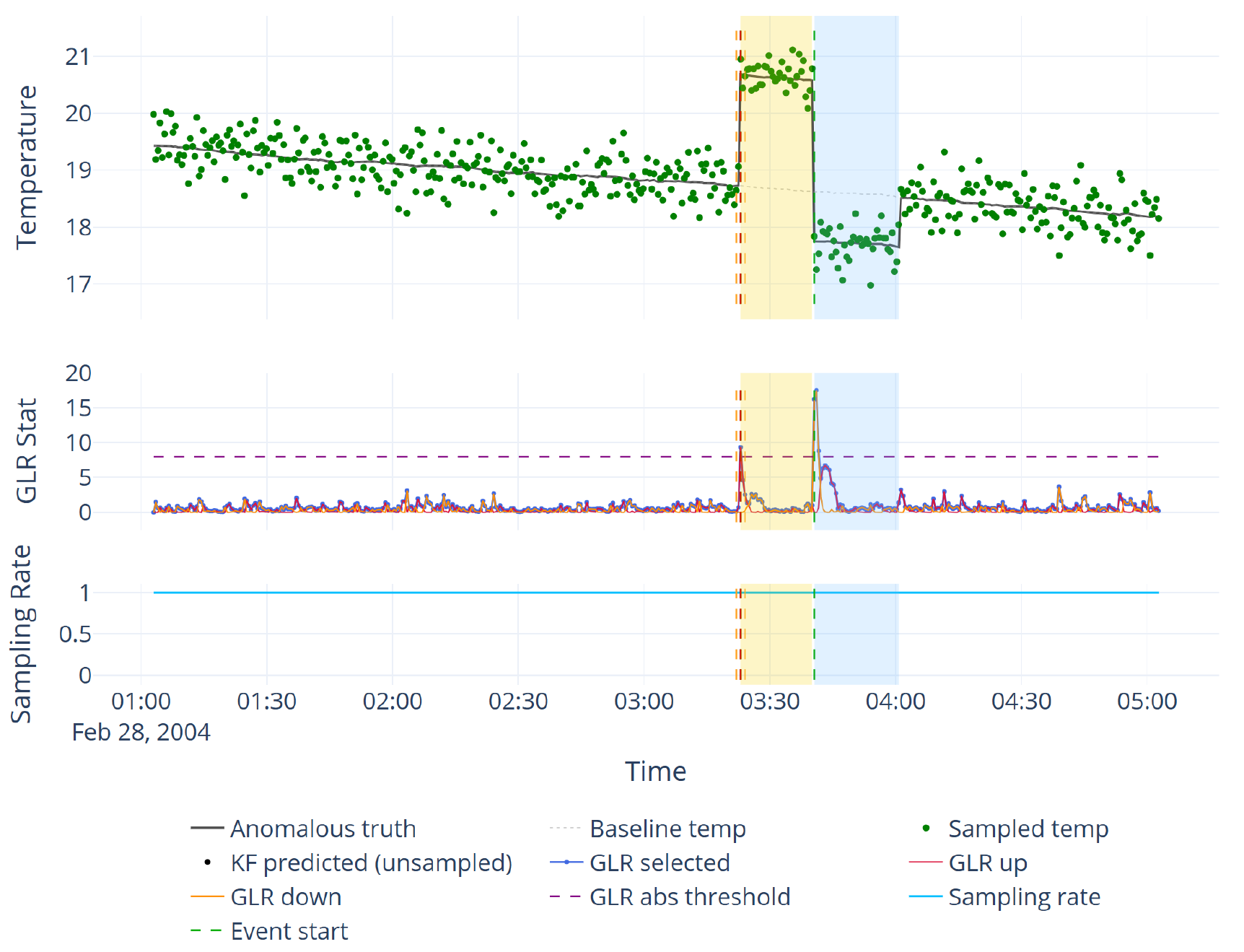}
\caption{Representative node-level diagnostics. Top: ordinary node 4 shows temporary alert-driven sampling densification within the anomaly window. Bottom: sentinel node 7 shows continuous sensing and a clear GLR onset peak.}
\label{fig:nodecase}
\end{figure}

Fig. \ref{fig:globaltemp} visualizes sentinel trajectories and anomaly windows in the representative run. It shows multiple localized events over time and confirms that continuously sampled sentinels capture clear changes near onset, supporting the gains in Coverage and AWSR reported in Table \ref{tab:task10main}. Fig. \ref{fig:nodecase} further illustrates the mechanism at node level: an ordinary node is temporarily densified inside an anomaly window, while a sentinel keeps continuous sampling and produces a clear GLR onset response.

\begin{table}[t]
\centering
\caption{Main results on Task 10.}
\label{tab:task10main}
\footnotesize
\setlength{\tabcolsep}{4.5pt}
\renewcommand{\arraystretch}{1.08}
\resizebox{\linewidth}{!}{
\begin{tabular}{lccccccc}
\toprule
\textbf{Method} & \textbf{EDR} & \textbf{ODD} & \textbf{Coverage} & \textbf{AWSR} & \textbf{GSR} & $\mathbf{RMSE_a}$ & \textbf{Energy} \\
\midrule
Full sampling          & 0.900 & 1.778 & 0.130 & 1.000 & 1.000 & 0.0000 & 392829 \\
Pure KF sparse         & 0.900 & 1.778 & 0.130 & 0.439 & 0.434 & 0.0040 & 170461 \\
Sentinel-assisted KF   & 0.900 & 1.444 & 0.977 & 0.933 & 0.758 & 0.0015 & 298127 \\
Oracle-local wake-up   & 1.000 & 0.000 & 1.000 & 1.000 & 0.856 & 0.0000 & 335447 \\
\bottomrule
\end{tabular}
}
\end{table}

\begin{figure}[t]
\centering
\includegraphics[width=0.9\linewidth]{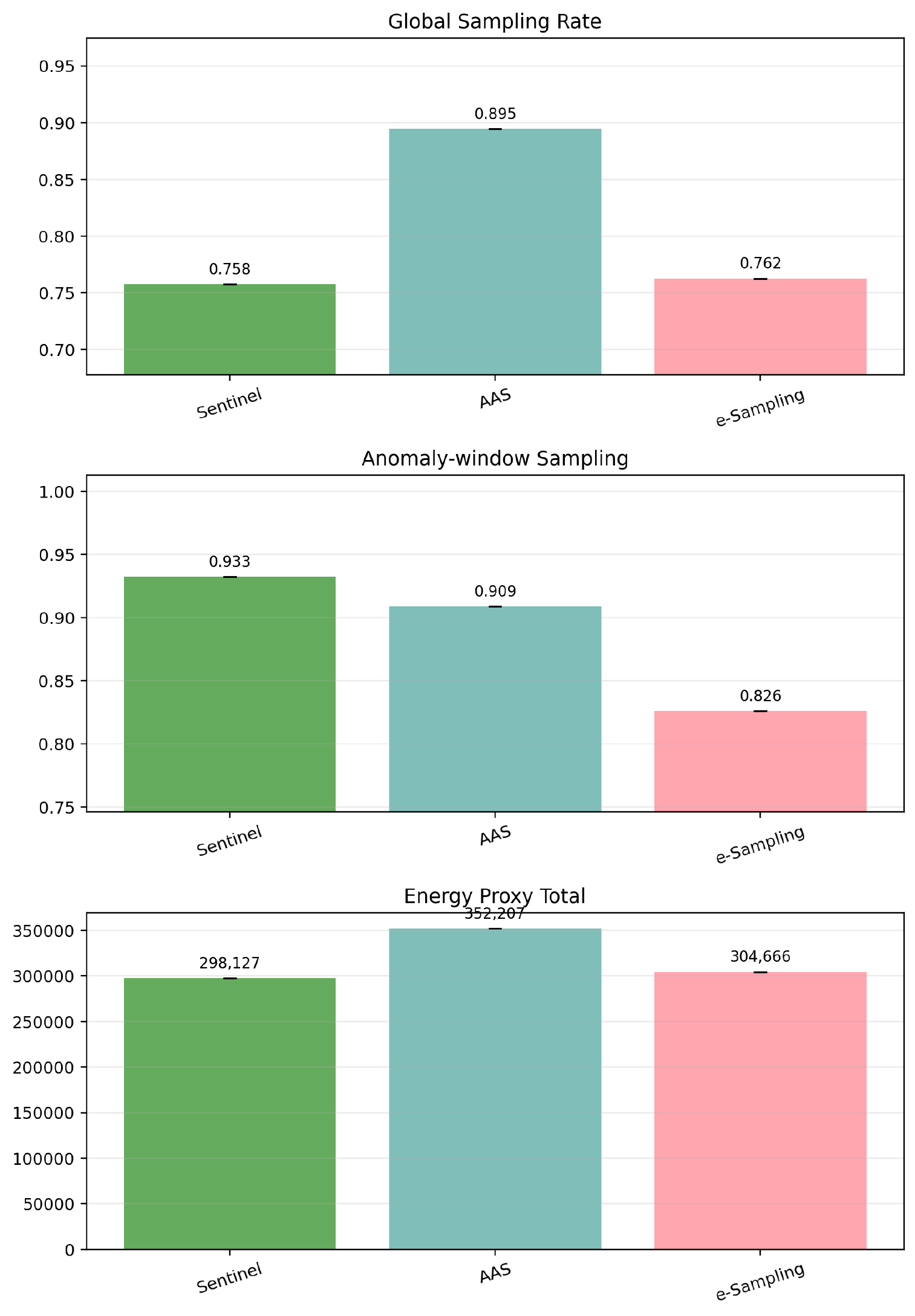}
\caption{Main cross-method results using GSR, AWSR, and Energy proxy total to show the sampling-cost trade-off relative to AAS and Adapted e-Sampling.}
\label{fig:comparemetric}
\end{figure}

\subsection{Experiment 2: Cross-Method Comparison}
\label{subsec:exp2}
Experiment 2 compares the proposed method with AAS and Adapted e-Sampling under a unified protocol, focusing on GSR, AWSR, and total energy proxy cost, with EDR and ODD reported as supplementary event-level indicators.

Fig. \ref{fig:comparemetric} reports the filtered cross-method comparison results. Relative to AAS, Sentinel-assisted KF reduces GSR from 0.895 to 0.758, improves AWSR from 0.909 to 0.933, and reduces the total energy proxy from 352267 to 298127. Relative to Adapted e-Sampling, it operates at nearly the same GSR (0.758 versus 0.762), but increases AWSR from 0.826 to 0.933 and reduces the total energy proxy from 304466 to 298127. These results support the system-level contribution: the method does not simply sample more everywhere, but reallocates sensing effort toward anomaly windows through sentinel-triggered local densification. In the supplementary event metrics, it also attains higher EDR and lower ODD than both baselines.

\subsection{Discussion}
\label{subsec:discussion}
By combining the results in Sections~\ref{subsec:exp1} and~\ref{subsec:exp2}, two main conclusions emerge. First, compared with Pure KF sparse sampling, the proposed method raises AWSR from 0.439 to 0.933 and Coverage from 0.130 to 0.977, showing that sentinel alerts can turn local detection into regional observation. Second, compared with Full sampling, it preserves a lower GSR and total energy proxy (0.758 and 298127 versus 1.000 and 392829), indicating that improved anomaly-window visibility is obtained without uniform full-rate sensing. Its main advantage over the retained external baselines is therefore a better trade-off between critical-period sampling and cost.

The evaluation is still limited by its use of the Intel Berkeley temperature dataset with injected abrupt spatiotemporal anomalies. This controlled setting enables fair comparison under a shared anomaly plan, data window, KF model, and random seed, but it does not capture sensor faults, environmental drift, overlapping disturbances, communication loss, or irregular field dynamics. Future validation should therefore test more random seeds, anomaly types, and deployment-like disturbances, and should incorporate real labelled WSN traces when available. Further work will also refine propagation control, recovery-aware exit, and node-level timing.

\section{Conclusion}
\label{sec:conclusion}

This paper presented a sentinel-assisted adaptive sampling framework for the tension between low-power operation and abrupt-event awareness in WSNs. By combining KF uncertainty-driven sparse sampling, hybrid GLR sentinel detection, one-hop alert propagation, recovery-aware alert control, and optimized sentinel placement, the method improves anomaly-window visibility while avoiding uniform full-rate sensing. Experiments show a better sampling-cost trade-off than the retained baselines. Future work will extend validation to more random seeds, richer anomaly scenarios, deployment-like disturbances, and real labelled WSN traces when available.

\bibliographystyle{IEEEtran}
\bibliography{ref}

\end{document}